\documentclass[aps,prl,twocolumn,superscriptaddress]{revtex4}
\usepackage{amssymb}
\usepackage{graphicx}

\newcommand{\ket}{\rangle}
\newcommand{\bra}{\langle}

\begin{document}
\bibliographystyle{apsrev}

\title{Density matrix renormalization group algorithms with a
single center site}
\author{ Steven R.\ White}
\affiliation{ 
Department of Physics and Astronomy,
University of California,
Irvine, CA 92697
}
\date{\today}
\begin{abstract}
\noindent 
We develop a correction to the density matrix used in density matrix renormalization group
calculations to take into account the incompleteness of the environment block. The
correction allows successful calculations using only a single site in the center
of the system, rather than the standard two sites, improving typical computation times
by a factor of two to four. In addition, in many cases where ordinary DMRG can get stuck
in metastable configurations, the correction eliminates the sticking. We test the
new method on the Heisenberg $S=1$ chain.

\end{abstract}
\pacs{PACS Numbers:  }
\maketitle

%\section{Introduction}

Since the density matrix renormalization group (DMRG) was
devloped\cite{dmrg,schollwock}, it has gradually been applied to more and more
difficult systems, such as wide ladders and 2D clusters, 
and systems with long-range interactions.
One of the
problems arising in these systems is the possibility that the simulation gets
stuck far from the ground state\cite{stripe}.  Several approaches have been developed to
overcome this problem, such as controlling the starting wavefunction
through potentials or quantum numbers, with the controls later removed.
Nevertheless, there has remained much room for improvement. In 1D
short-range systems, the standard DMRG finite system algorithm avoids
convergence problems remarkably well because of the presence of the second
center site in the block configuration. However, the extra site increases
the computation time and memory requirements.  An alternative to utilizing
the extra site, which works better in the more difficult cases, has not
been available. In this paper we describe such an alternative method, which
relies on a correction to the reduced density matrix in order to retain a
broader variety of states.

In the top panel of  Fig. 1 we show the ``superblock'' configuration for the
standard finite-system algorithm, where
the lattice is divided into two large blocks, the system and the envirionment blocks,
both with truncated bases, 
with two sites between them.
The algorithm for a single DMRG step consists of
finding the ground state for this ``superblock''; obtaining the density
matrix for the system block plus site; diagonalizing this density matrix; and
then changing basis to the most probable eigenvectors of the density matrix.
This step replaces the system block, described by $m$ states, by a block
one site larger, but also described (approximately) by $m$ states.
One then shifts the dividing line between the system and environment by one site,
in order to add another site to the system block,
and repeats the process. When the system block encompasses the whole system,
the direction is reversed and the roles of system and environment blocks are
reversed. A sweep consists of one pass back and forth through the system. In a
simple 1D spin system one often obtains convergence to very high accuracy, e.g. an
accuracy in the energy of order $10^{-10}$, with one
or two sweeps through the lattice.

\begin{figure}[tb]
\includegraphics*[width=4cm]{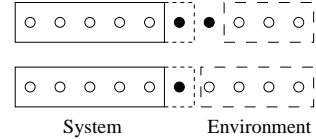}
\caption{
Standard two-site DMRG method (top) and the single site method.
}
\label{figone}
\end{figure}

In this description, it is apparent that one of the two sites in the center is
crucial to the algorithm.  The role of the other site
is to increase the dimension and also the accuracy of the environment, particularly
at the point where it connects to the system.
One can leave out this extra site, i.e. use an environment block with the site
already a part of it, as shown in the bottom panel of Fig. 1.
This decreases the computation time for a step by roughly the
number of states in a single site. However, one finds that even in 1D systems,
the progress towards the ground state is much slower, and often stops
altogether far from
the true ground state. This can be understood in various ways. For example,
suppose the ground state has total $z$ component of spin $S_z=0$, and also suppose the 
environment block is poor and only has states with $S_z=0$. 
Then the renormalized system block will only have states with $S_z=0$, and no
fluctuations in the spin will develop between the two blocks. In fact, any
limitation on the quantum numbers present in the environment translates into
a restriction on the states appearing in the renormalized system block. The distribution
of states between various quantum numbers in the environment also translates directly
to the renormalized system block.  Note that if the environment block has
$m$ states, then the maximum
number of nonzero eigenvalues of the density matrix is also $m$, and the
number of states never increases unless states are added ``artificially'' despite
having density matrix eigenvalues of zero. Simple fixes, such as adding extra
random states with a larger range of quantum numbers, improve but do not fix the
very poor convergence of the single site algorithm. 

The essential problem here arises when a particular fluctuation between
the system and environment which
should be present is not because the environment block
does not have the relevant states. Hence, the fluctuation is not represented
in the density matrix and the new system block will not possess its relevant
states for that fluctuation. Later, when the roles of system and environment are
reversed, the relevant states again do not appear. In a 1D system with short
range interactions the extra environment site
does a very good job  of ensuring that the most relevant 
fluctuations are at least approximately present the environment, so that
subsequent sweeps can build in the fluctuations to high accuracy.

In wide ladders or systems with longer
range interactions, the addition of a single site to the environment is
not always adequate. There may be missing fluctuations which are far from
the extra site, and so are never built in. Even in these cases the extra
site allows $m$ to increase sensibly and as one lets $m \to \infty$ one
obtains exact results. However, for practical values of $m$ one may find
unacceptably slow convergence.

In this paper we describe an approximate correction to the density matrix
to describe the key states which have been left out because the environment block
is inadequate. With this correction, the single site superblock configuration 
converges well. In addition, convergence in more difficult systems is
dramatically improved, in either the single site or two site configurations.
We present two different derivations of the correction, and give examples
using the $S=1$ Heisenberg chain.

We first give a simple, rough argument. Consider the power method for finding
the ground state: iterate $\psi_{n+1} = (1 - \varepsilon H) \psi_n$, where $\varepsilon$ 
is a small constant. As long as $\psi_0$ is not orthogonal to the exact
ground state, and $\varepsilon$ is small enough, the power method is
guaranteed to converge to the ground state. Consequently, if the basis represents
both $\psi$ and $H \psi$ exactly, and we minimize the energy
within this basis, we expect exact convergence. The crucial point is the need
to enlarge the basis to represent $H \psi$. Within the standard DMRG basis obtained
from $\psi$, 
after solving for the ground state, $H \psi = E \psi$, and nothing is changed by
adding $H \psi$ to the basis. To go beyond the basis, we need to construct
the parts of $H \psi$ as the basis is built up. The crucial 
terms of $H \psi$ come from the terms of $H$ which connect the system and
environment blocks. 

For the current superblock configuration, write the Hamiltonian in the form 
\begin{eqnarray}
H = \sum_\alpha t_\alpha \hat A^\alpha \hat B^\alpha.
\end{eqnarray}
Here the $\hat A^\alpha$ act only on the system block (including the site to be added to it),
and the 
$\hat B^\alpha$ act only on the environment block (plus its site). 
All the terms which do not connect the blocks are contained in two terms of
the sum which have either
$A$ or $B$ equal to the identity operator, so that this form is
completely general. (The other term in each case is the block Hamiltonian.)
In order to put $H \psi$ into the basis, we need to target, in addition to
$\psi$, the terms ${\hat A^\alpha \psi}$ for all $\alpha$. Let the states of the
system have indices $s$, $p$, and $q$, and the states of the environment $e$. The
state $\hat A^\alpha \psi$ can be written as
\begin{eqnarray}
\sum_{se} \sum_{p} A^\alpha_{sp} \psi_{pe} | s\ket| e\ket.
\end{eqnarray}
Targetting this wavefunction  means adding into the density matrix a term
\begin{eqnarray}
\Delta \rho^\alpha_{ss'} = a_\alpha \sum_{epq} A^\alpha_{sp} \psi_{pe} \psi_{qe}^* {A^\alpha_{s'q}}^*
\end{eqnarray}
where $a_\alpha$ is an arbitrary constant determining how much weight to put into
this additional state.
The total contribution of all the terms is
\begin{eqnarray}
\Delta \rho = \sum_{\alpha} a_\alpha \hat A^\alpha \rho \hat A^{\alpha\dagger}
\end{eqnarray}
where $\rho$ is the density matrix determined in the usual way, only from $\psi$.
This is the form of the correction that we use, with $a_\alpha = a \sim 10^{-3}-10^{-4}$.

As a second derivation, we utilize perturbation theory.
First, imagine that the environment block, but not the
system block, is complete.
We obtain the ground state exactly for this
superblock, and then transform to the basis of density
matrix eigenstates for the system block, and then also do the
same for the environment block. Then the wavefunction can be written
in the form
\begin{eqnarray}
|\psi \ket= \sum_s \psi_s |L_s\ket |R_s\ket .
\label{psieqn}
\end{eqnarray}
The reduced density matrix is
\begin{eqnarray}
\label{rhoeqn}
\rho = \sum_s\bra R_s|\psi \ket \bra \psi | R_s \ket
= \sum_s |\psi_s|^2 |L_s\ket \bra L_s|
\end{eqnarray}

Now consider the realistic case where the environment block is not
complete. Assume the incompleteness takes the simple form that some of 
the $|R_s\ket$ are missing, labeled $\bar s$, whereas $s$ are present.
Let $P$ be a projection operator for the environment block $P = \sum_s |s\ket\bra s|$, 
and take $\bar P=1-P$.   
Let the unperturbed ground state, with energy $E_0$ and density matrix $\rho_0$, 
be obtained using the incomplete environment basis.
We take as a perturbation the terms in the Hamiltonian
which couple to the states $\bar s$, namely
\begin{eqnarray}
H' = \sum_\alpha t_\alpha \hat A^\alpha (\bar P \hat B^\alpha P + P \hat B^\alpha \bar P).
\end{eqnarray}
The first order perturbative correction to the wavefunction due to $H'$ is
\begin{eqnarray}
\label{psipeqn}
|\psi' \ket= \sum_\alpha t_\alpha (E_0-H_0)^{-1}
\hat A^\alpha \bar P \hat B^\alpha |\psi \ket
\end{eqnarray}
where $H_0 = H-H'$. 

In order to make progress we assume
that each perturbation term $A^\alpha \bar P \hat B^\alpha$ acting on the
ground state creates a set of nearly degenerate excited states,
with average energy $E_\alpha$. This assumption is equivalent to
saying that the spectral function associated with each term is dominated
by a narrow peak at $E_\alpha$.  This significant
approximation is reasonable because the correction to the density matrix
is only used to enlarge the basis, to improve DMRG convergence. 
Correspondingly, we approximate
$(E_0-H_0)^{-1}$ as $(E_0-E_\alpha)^{-1} \equiv 1/\varepsilon_\alpha$.
This gives
\begin{eqnarray}
\label{psipeqntwo}
|\psi' \ket\approx \sum_s \psi_s \sum_\alpha \frac{t_\alpha }{\varepsilon_\alpha}
\hat A^\alpha \bar P \hat B^\alpha |L_s\ket |R_s\ket .
\end{eqnarray}
There are no first order corrections to the density matrix from
$|\psi'\ket$, since $\bar P |\psi \ket = 0$. The lowest order correction
to $\rho$ can be written as
\begin{eqnarray}
\label{rhoeqntwo}
\Delta \rho = \sum_{ss'} \psi_s \psi^*_{s'} \sum_{\alpha{\alpha'}} 
\frac{t_\alpha}{\varepsilon_\alpha}
\frac{t_{\alpha'}}{\varepsilon_{\alpha'}}
\hat A^\alpha |L_s\ket \bra L_{s'}| \hat A^{\alpha \dagger}
M_{s'\alpha'\alpha s}
\end{eqnarray}
where
\begin{eqnarray}
\label{rhoeqnthree}
M_{s'\alpha'\alpha s} = 
\bra R_{s'} | \hat B^{\alpha'\dagger}  \bar P \hat B^\alpha  |R_s\ket 
\end{eqnarray}

Here if $A$ is the unit operator, the term adds nothing to the basis. 
If $B$ is the unit operator, $M_{s'\alpha'\alpha s}$ vanishes.
For the nontrivial pairs of operators $A$ and $B$, this matrix
element somewhat
resembles a correlation function and it is natural to assume that the
diagonal terms are dominant, where $\alpha={\alpha'}$ and $s=s'$. 
We expect the off diagonal terms $\alpha\ne{\alpha'}$ to describe coherence
between different perturbation terms which would tend to reduce the number of
basis functions needed to describe the system block; therefore, ignoring 
the offdiagonal terms is a conservative assumption.
Accordingly, we
take
\begin{eqnarray}
M_{s'\alpha'\alpha s}
\approx \delta_{s{s'}} \delta_{\alpha{\alpha'}} b_\alpha
\end{eqnarray}
This gives Eq. (4) with
$a_\alpha = b_\alpha |t_\alpha|^2/\varepsilon_\alpha^2$, and where
we omit block-Hamiltonian terms. 

In practice, we take $a_\alpha$ to be a small constant $a$ independent of $\alpha$.
Construction of the correction to $\rho$ take a calculation time for a single
step proportional to $m^3$ times the number of connecting terms, which is typically
significantly smaller than the other parts of the DMRG calculation, although the
scaling is the same. Larger values of $a$ introduce more ``noise'' into the basis,
speeding convergence, but also limiting the final accuracy. 
Note that it is just as easy to apply the correction within the two-site method 
as the single-site
method, which may be useful in some very difficult cases. We do not present results
for this combination here.

\begin{figure}[tb]
\includegraphics*[width=5cm]{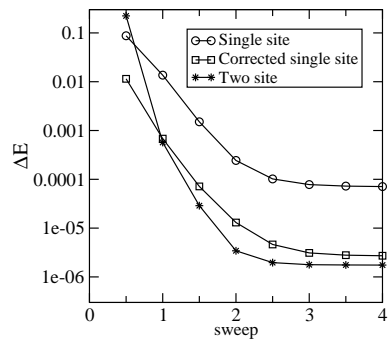}
\caption{
Error in the total energy for a 100 site Heisenberg spin-one chain, keeping
$m=50$ states per block, and using open
boundaries terminated with $S=1/2$ spins to remove the $S=1/2$ end 
states (98 $S=1$ sites + 2 $S=1/2$'s).
The results are displayed for each half-sweep corresponding to reaching either
the left or right end of the system. 
The two site method is the standard DMRG approach. 
The numerically exact energy was determined
with the two-site method, using $m=200$.
}
\label{figtwo}
\end{figure}

As a first test calculation, we consider the $S=1$ Heisenberg model
\begin{equation}
H = \sum_j \vec{S}_j \vec{S}_{j+1} \;,
\end{equation}
where we have set the exchange coupling $J$ to unity.
The correction consist of the following: for each boundary site $i$ 
of a block, i.e. a site
directly connected to the other block, we add into the density matrix
\begin{eqnarray}
\label{rhopeqn}
\Delta \rho = a (S^+_i \rho S^-_i + S^-_i \rho S^+_i + S^z_i \rho S^z_i).
\end{eqnarray}
For a chain with open boundaries, there is one site $i$; for periodic boundaries,
there are two.
One could argue that this expression should be adjusted with factors of 2 between
the $z$ term and the other two terms, but this is not likely to make a significant
difference.
Note that the $S^+$, $S^-$ terms automatically increase the
range of quantum numbers (i.e. total $S^z$) with nonzero density matrix eigenvalues.
Figure 2 shows the convergence of the energy for a 100 site chain 
with open boundaries as a function of the sweep, keeping
$m=50$ states, relative to the numerically exact
result obtained with $m=200$ and 10 sweeps.  One can see the excellent convergence
of the standard approach. The single-site method without corrections does not
do too badly in this case, but still gets stuck significantly above the two-site
energy. Adding the corrections, in this case with $a=10^{-4}$, dramatically
improves the convergence, making the single site method converge nearly as fast as
the two site method. The two site method is roughly a factor of 
three slower than the single site method. Thus, even in this simple 1D case where the standard
approach works extremely well, there are advantages to using the corrected
single site method.

\begin{figure}[tb]
\includegraphics*[width=5cm]{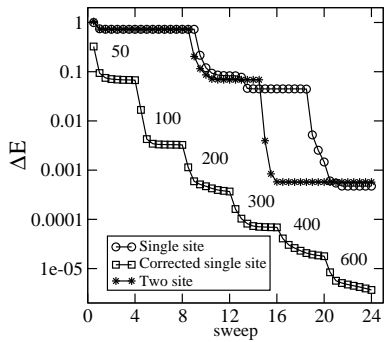}
\caption{
Error in the total energy for a 100 site Heisenberg spin-one chain, with periodic
boundaries.
The number of states kept per block is indicated, and is the same for all three
methods; four sweeps were made for each $m$.
The correction parameter $a$ was taken to be $10^{-4}$ for sweeps 1 - 8, and
$10^{-6}$ for later sweeps. A somewhat slower convergence is visible for
$a=10^{-6}$. 
The reference energy used was 100 times the infinite energy per site,
-1.401484038971(4)\cite{whitehuse}. The corrected single site method using $m=4000$
states gives a slightly lower total energy, due to exponentially small finite size effects,
of -140.14840390392.
}
\label{figthree}
\end{figure}

The results change significantly if we consider periodic boundary conditions.
Here we consider the same superblock configuration as with open boundary
conditions, but simply add in the connection to the Hamiltonian between the 
first and last sites. There are better configurations for periodic boundaries,
such as considering it to be a ladder with the interchain couplings turned off
except at the ends. These other configurations are superior only in the sense
of improved convergence with the number of sweeps, not improved with respect
to the number of states for a large number of sweeps. 
This naive configuration thus provides a difficult test for the single site method
with corrections.
In Fig. 3, we show the results for the same three cases as in Fig. 2. In this case,
in the early sweeps, both uncorrected methods are stuck, ignoring the extra link
between the first and last sites. The extra link eventually appears in the basis,
but there is still sticking two or three times in higher energy states. In contrast,
the corrected single site method never gets stuck and shows excellent convergence.

A very useful DMRG technique is the extrapolation of the energy with
the truncation error, i.e. the weight in the states which are thrown out.
If the truncation error were measured exactly, with a
complete basis for the environment, then the energy error would
be proportional to the truncation error, allowing a linear extrapolation to
zero truncation error. In practice, the apparent truncation error from the
two site method may often be an underestimate,
but one often finds that it is very consistent and still allows
excellent extrapolation, even on fairly wide ladders. 
The truncation error within the corrected single-site
method depends on $a$: as $a \to 0$, the apparent truncation error goes to zero
and is unrelated to the exact truncation error. However, if $a$ is not too small,
linearity and excellent extrapolation are possible. 

Figure 4 shows results for the 100 site periodic system with a larger value of $a$,
$10^{-2}$, suitable for extrapolation. The results show excellent linearity.
The extrapolation gives -140.148416, off by $1.2\times10^{-5}$, whereas the
sweep with $m=340$ gave -140.148279, off by $1.2\times10^{-4}$. We have 
found that typically an order of magnitude improvement in the estimate for the energy is
obtained by extrapolation in good cases; here we see similar improvement.
In performing these extrapolations one always need to check the linearity
for the system being studied.

\begin{figure}[tb]
\includegraphics*[width=5cm]{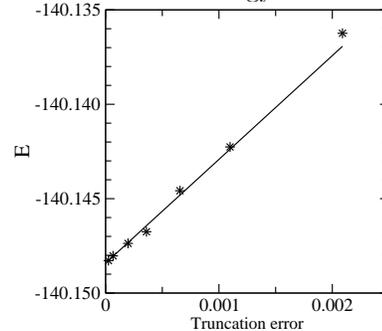}
\caption{
Error in the total energy for the system of Fig. 3 versus the truncation error,
with $a=10^{-2}$. In this run two sweeps for each value of $m$, were made. The
points shown are for $m = $80, 100, 120, 160, 200, 260, and 340.
The line is a linear extrapolation, weighted with a standard deviation for each
point assumed to be proportional to the truncation error at that point.
}
\label{figfour}
\end{figure}

In summary, we have demonstrated a correction to the density matrix which allows
the single-site DMRG method to converge well, and which improves the convergence 
dramatically for hard-to-converge systems.

We thank J. Rissler and F. Verstraete for helpful conversations.
We acknowledge the support of the NSF under grant DMR03-11843.

%----------------------------------------------------------------------
{}
%-------------------------------------------------------

\end{document}